\documentclass[preprint,twocolumn,3p,nopreprintline,sort&compress]{elsarticle}
\usepackage{epsfig}
\usepackage{amsbsy,amssymb,amsmath,amsthm} 
\usepackage{graphicx}

\usepackage{array}

\usepackage{graphicx}
\usepackage{dcolumn}
\usepackage{bm}

\def\db#1{ D_{#1}}

\def\slh#1{\rlap / {#1}}

\def\C++{{{\sc c++ }}}
\def\zeta{{\mathbf{z}}}
\def\idea{{\Gamma}}
\def\emm{{m}}

\def\spa#1.#2{\langle#1\,#2\rangle}
\def\spb#1.#2{[#1\,#2]}

\def\spab#1.#2.#3{\langle\mskip-1mu{#1}
                  | #2 | {#3}]}

\def\spba#1.#2.#3{[\mskip-1mu{#1}
                  | #2 | {#3}\rangle}

\def\spbb#1.#2.#3.#4{[\mskip-1mu{#1}
                     | {#2} \ {#3} | {#4}]}

\def\spaa#1.#2.#3.#4{\langle\mskip-1mu{#1}
                     | {#2} \ {#3} | {#4}\rangle}

\newcommand{\beq}{\begin{equation}}

\newcommand{\eeq}{\end{equation}}

\newcommand{\bea}{\begin{eqnarray}}

\newcommand{\eea}{\end{eqnarray}}

\newcommand{\bean}{\begin{eqnarray*}}

\newcommand{\eean}{\end{eqnarray*}}
           
\newcommand{\nn}{\nonumber \\}

\newtheorem{prop}{Proposition}[section]

\newtheorem{theo}{Theorem}[section]

\newcommand{\mpi}{Max-Planck-Institut f\"ur Physik, F\"ohringer Ring 6, 80805 M\"unchen, Germany}
\newcommand{\padova}{Dipartimento di Fisica e Astronomia, Universit\`a di Padova, and INFN                                    
Sezione di Padova, via Marzolo 8, 35131 Padova, Italy}
\newcommand{\cuny}{New York City College of Technology, City University of New York, 300 Jay Street, Brooklyn, NY 11201, USA}
\newcommand{\cunygc}{The Graduate School and University Center, City University of New York, 365 Fifth Avenue, New York, NY 10016, USA}

\begin{document}


\begin{frontmatter}

\title{Scattering Amplitudes from Multivariate Polynomial Division}

\author{Pierpaolo  Mastrolia}
\ead{pierpaolo.mastrolia@cern.ch}
\address{\mpi}
\address{\padova}

\author{Edoardo Mirabella}
\ead{mirabell@mppmu.mpg.de}  
\address{\mpi}

\author{Giovanni Ossola} 
\ead{GOssola@citytech.cuny.edu}
\address{\cuny}
\address{\cunygc}

\author{Tiziano Peraro} 
\ead{peraro@mppmu.mpg.de}
\address{\mpi}

\begin{abstract}
We show that the evaluation of scattering amplitudes can be
formulated  as a problem of multivariate polynomial division, with
the components of the integration-momenta as indeterminates.
We present a recurrence relation which, independently of the
number of loops, leads to
the multi-particle pole decomposition of  the integrands of the
scattering amplitudes.
The recursive algorithm is based on the Weak Nullstellensatz Theorem
and on the division modulo the Gr\"obner basis associated to all
possible multi-particle cuts.
We apply it to dimensionally regulated one-loop amplitudes,
recovering the well-known integrand-decomposition formula.
Finally, we focus on the maximum-cut,
defined as a system of on-shell conditions constraining the components
of all the integration-momenta. 
By means of the Finiteness Theorem and of the Shape Lemma, 
we prove that the residue at the maximum-cut is parametrized 
by a number of coefficients equal to the number of solutions of the cut itself.
\end{abstract}



\end{frontmatter}


\section{Introduction}
 
Scattering amplitudes in quantum field theories are analytic
functions of the momenta of the particles involved in the scattering
process, and can be determined by their singularity structure. 
The multi-particle factorization properties of the amplitudes are
exposed when propagating particles go on their mass-shell
\cite{Bern:1994zx,Britto:2004nc,Cachazo:2004kj,Britto:2004ap}.

The investigation of the residues at the
singular points has been fundamental for 
discovering new relations fulfilled by scattering amplitudes. 
The BCFW  recurrence relation \cite{Britto:2004ap},
its link to the leading singularity of one-loop amplitudes \cite{Britto:2004nc},
and the OPP integrand-decomposition formula for  one-loop
integrals \cite{Ossola:2006us} have shown the underlying simplicity beneath 
the rich mathematical structure of quantum field theory.
Moreover they have become efficient techniques leading to  quantitative predictions 
at the next-to-leading order in
perturbation theory \cite{Berger:2008sj,Giele:2008bc,Badger:2010nx,Bevilacqua:2011xh, Hirschi:2011pa, Cullen:2011ac,Agrawal:2011tm,Cascioli:2011va}.

The integrand reduction methods  \cite{Ossola:2006us} allow to 
decompose one-loop amplitudes in terms of Master Integrals (MI's) 
without performing the loop integration,
and are based on the multi-particle pole expansion of the integrand. 
The expansion is equivalent to
the decomposition of the numerator in terms of (a combination
of) products of  denominators,  with polynomial coefficients.
In the context of an integrand-reduction, any  integration 
is replaced by {\it polynomial fitting}.

The first extension of the {\it integrand reduction method} beyond one-loop
was proposed in \cite{Mastrolia:2011pr}, and it was used to reproduce
the results of two-loop 5-point planar and non-planar amplitudes in
${\cal N}=4$ SYM \cite{Bern:2006ew,Carrasco:2011mn}.
A key point of the higher-loop extension is the proper parametrization
of the residues at the multi-particle poles.
Each residue is a multivariate polynomial in the 
{\it irreducible scalar products} (ISP's) among the loop momenta and 
either external momenta or polarization vectors constructed out of
them.  ISP's cannot be expressed in terms of denominators, thus 
any monomial formed by ISP's is the numerator of a potential MI which 
may appear in the final result.
Hence, a systematic classification of the polynomial structures of the residues is mandatory.
In~\cite{Mastrolia:2011pr}, the residues have been obtained by relating the 
ISPs to monomials in the components of the loop momenta expressed
in a basis chosen according to the topology of the on-shell diagram.

Badger, Frellesvig and Zhang \cite{Badger:2012dp} combined
on-shell conditions with Gram-identities \cite{Gluza:2010ws} to limit
the number of monomials appearing in the residues.  This technique was
applied to the integrand decomposition of two-loop 4-point planar and
non-planar diagrams in supersymmetric as well as non-supersymmetric YM theories.

In this work, we show that the shape of the residues is uniquely determined 
by the on-shell conditions alone, without any additional constraint. 
We derive a simple {\it integrand recurrence relation} 
that generates the required multi-particle pole decomposition for arbitrary amplitudes, 
independently of the number of loops. 

The  algorithm treats   the numerator and the denominators of any Feynman integrand, 
as multivariate polynomials in the components of the loop variables. 
The properties of multivariate polynomials have been extensively studied in the mathematical
literature, see {\it e.g.}~\cite{buch1,Cox1,Cox2,moller,rouillier,sturmfels,verschelde}. 
The method uses  both the {\it weak Nullstellensatz theorem} 
and the {\it multivariate polynomial division} modulo 
appropriate Gr\"obner basis~\cite{buch1}. 
In the context  of
the integrand reduction, 
univariate polynomial division  has been already  introduced in \cite{Mastrolia:2012bu}
to improve  the decomposition of one-loop scattering 
amplitudes.

\medskip

The algorithm, which is  described in Section II, relies on 
 general properties of the loop integrand:
\begin{itemize}
\item When the number $n$ of denominators is larger than the total number of the
components of the loop momenta,
the {\it weak Nullstellensatz theorem} yields the trivial reduction of
an $n$-denominator integrand in terms of integrands with $(n-1)$ denominators.
\item When  $n$ is equal or less than the total number of
components of the loop momenta, we divide the numerator modulo the
Gr\"obner basis of the $n$-ple cut, namely modulo a set of polynomials
vanishing  on the same on-shell solutions as the cut denominators. The {\it remainder}
of the division is the {\it residue}  of the $n$-ple cut.  The {\it quotients} 
generate integrands with $(n-1)$ denominators which should undergo the same decomposition.
\item By iterating this procedure,  we extract  
the polynomial forms of {\it all} residues.
The algorithm will stop when all cuts are exhausted, and no denominator is left,
leaving us with the integrand reduction formula.
\end{itemize}

\medskip

In Section III  we apply the algorithm  to a generic one-loop
integrand, reproducing  the $d$-dimensional integrand decomposition formula 
\cite{Ossola:2006us,Ellis:2007br,Giele:2008ve,Ellis:2008ir}.

\medskip

In Section IV we conclude by  proving a theorem on the {\it maximum-cuts}, i.e. 
the cuts defined by the maximum number of on-shell conditions which can
be simultaneously satisfied by the loop momenta.   The on-shell conditions 
of a maximum cut lead to a zero-dimensional system. 
The 
{\it Finiteness Theorem} and the 
{\it Shape Lemma} ensure that the
residue at the maximum-cut is parametrized 
by $n_s$ coefficients, where $n_s$ is the number of
solutions of the multiple cut-conditions.    
This guarantees that the
corresponding residue can always be reconstructed by evaluating the numerator
at the solutions of the cut.
 \\

During the completion of this work, Zhang has presented an algorithm \cite{Zhang:2012ce}
embedding the ideas presented in \cite{Badger:2012dp} within more general techniques of 
algebraic geometry, among which the division modulo Gr\"obner basis is used as well.

 \section{Multivariate  polynomial division}
 \label{Sec:Gen}

In what follows, we assume 4-dimensional loop-momenta.
Extensions to higher-dimensional cases, according to the chosen
dimensional regularization scheme, can be treated analogously - as we
will show when discussing the one-loop integrand reduction.

 The integrand reduction methods~\cite{Ossola:2006us,Ossola:2007bb,Ossola:2007ax,Ossola:2008xq,Ellis:2007br,Giele:2008ve,Ellis:2008ir,Mastrolia:2008jb, Mastrolia:2012bu,Mastrolia:2011pr,Badger:2012dp} recast the problem of computing 
 $\ell$-loop amplitudes with $n$ denominators as the reconstruction  of  integrand functions of the type
 \beq
 \mathcal{I}_{i_1\cdots i_n} \equiv 
\frac{ {\cal N}_{i_1\cdots i_n} ( q_1, \ldots , q_\ell   )}{\db{i_1}(q_1, \ldots ,q_\ell ) \cdots \db{i_n}(q_1, \ldots  ,q_\ell )} \; ,
 \label{Eq:IntegrandM}
\eeq
where $q_1, \ldots , q_\ell$ are integration momenta.
The generic propagator can be written as follows:
 \beq
\db{i} = \left ( \sum_{j=1}^\ell \alpha_j \; q_j   + p_i \right )^2 - m_i^2 \; , \,
\alpha_j \in \{ 0,\pm 1  \}  \; . 
\eeq
The numerator ${\cal N}_{i_1\cdots i_n}$ and any of the denominators $\db{i}$ 
are polynomial in  the components of the loop momenta, 
 say $\zeta \equiv (z_1, \ldots,  z_{4\ell})$, {\it i.e.} 
 \beq
 \mathcal{I}_{i_1\cdots i_n}  = \frac{{\cal N}_{i_1\cdots i_n}(\zeta)}{D_{i_1}(\zeta) \cdots D_{i_n}(\zeta)} \; . 
 \label{Eq:Igen}
 \eeq

Let us consider the  ideal generated by the $n$ denominators in Eq.~(\ref{Eq:Igen}) ,
 \bea
{\cal J}_{i_1 \cdots  i_n} &=&  \langle D_{i_1}, \cdots , D_{i_n} \rangle  \nn
&\equiv& \left \{\sum_{\kappa=1}^n  h_{\kappa}(\zeta)  D_{i_\kappa}(\zeta) :  h_\kappa(\zeta) \in P[\zeta] \right \} ,
\nonumber 
\eea
 where $P[\zeta]$ is the set of polynomials in $\zeta$.
 The common zeros of the elements of 
${\cal J}_{i_1 \cdots  i_n} $ are exactly the common zeros of the denominators.

The multi-pole decomposition  of Eq.~(\ref{Eq:IntegrandM}) is explicitly 
achieved by performing multivariate polynomial division,
yielding an expression of ${\cal N}_{i_1\cdots i_n}$
in terms of denominators and residues.\\
We construct a Gr\"obner basis~\cite{buch1} 
(see Ch. 2 of \cite{Cox1}), generating the ideal ${\cal J}_{i_1 \cdots i_n}$
with respect to a chosen monomial order,
\beq
\mathcal{G}_{i_1 \cdots  i_n}=\{g_{1}(\zeta), \ldots  , g_{\emm}(\zeta) \}  \;  . 
\eeq
Unless otherwise indicated, we will assume lexicographic order.

In this formalism, the $n$-ple cut-conditions
$\db{i_1} = \ldots =\db{i_n} = 0$, 
are equivalent to $g_1 = \ldots = g_\emm = 0$. 
The number $\emm$ of elements of the Gr\"obner basis is the 
{\it  cardinality} of the basis. In general, $m$ is different from $n$.
We then consider the multivariate division of ${\cal N}_{i_1\cdots i_n}$ modulo 
$\mathcal{G}_{i_1\cdots i_n}$ (see Ch. 2, Thm. 3 of \cite{Cox1}),
\beq
{\cal N}_{i_1\cdots i_n}(\zeta) = \idea_{i_1 \cdots  i_n}  + \Delta_{i_1\cdots i_n}(\zeta)  \; ,
\label{Eq:DecGenI}
\eeq
where $\idea_{i_1 \cdots  i_n} = \sum_{i=1}^\emm  \mathcal{Q}_{i}(\zeta)
g_i(\zeta) $ 
is a compact notation for the sum of the products of the quotients $\mathcal{Q}_{i}$ 
and the divisors $g_i$.
The polynomial  $\Delta_{i_1 \cdots i_n}$ is the remainder of the
division. Since ${\cal G}_{i_1 \cdots  i_n}$ is a Gr\"obner basis, 
the remainder is uniquely determined once the monomial order is fixed.\\ 
The term $\idea_{i_1 \cdots  i_n}$ belongs to the ideal
${\cal J}_{i_1 \cdots i_n}$,  thus it can be expressed 
in terms of denominators, as
\beq
\idea_{i_1 \cdots  i_n}=  \sum_{\kappa=1}^{n}   
{\cal N}_{i_1\cdots i_{\kappa -1}i_{\kappa+1}\cdots i_n}(\zeta) \db{i_\kappa}(\zeta) \; . 
\label{Eq:fromGtoD}
\eeq  
The explicit form of 
${\cal N}_{i_1\cdots i_{\kappa -1}i_{\kappa+1}\cdots  i_n}$ can be
found by expressing the elements of the Gr\"obner basis 
in terms of the denominators.

\subsection{Reducibility criterion.}  
An integrand  $\mathcal{I}_{i_1\cdots i_n} $
 is  said to be reducible if it  
 can be written in terms of lower-point integrands: that happens when 
the numerator can be written in terms of denominators.
The concept of {\it reducibility} can be formalized in algebraic geometry.
Indeed a  direct consequence of Eqs.~(\ref{Eq:DecGenI}) and (\ref{Eq:fromGtoD})
is the following

\begin{prop}
 The   integrand  $\mathcal{I}_{i_1\cdots i_n} $
 is reducible iff 
 the remainder of the division modulo a Gr\"obner basis vanishes, i.e. iff 
${\cal   N}_{i_1\cdots i_n} \in \mathcal{J}_{i_1\cdots i_n}$. 
\label{Prop:Red}
\end{prop}

\noindent 
Proposition~\ref{Prop:Red} allows to prove 

 \begin{prop}
 An integrand ${\cal I}_{i_1, \ldots, i_n}$ is reducible if the cut $(i_1, \ldots, i_n)$ leads to a system of equations with no solution. 
 \label{Prop:Red4D}
 \end{prop}
\begin{proof}
In this case, the system is over-constrained.
The $n$ propagators cannot vanish simultaneously, i.e.
\beq
D_{i_1} (\zeta) = \cdots = D_{i_n}(\zeta)  =0 
\eeq 
has no solution.  Therefore, according to the {\it weak Nullstellensatz} theorem 
(Thm. 1, Ch. 4 of \cite{Cox1}),
\beq
1 = \sum_{\kappa=1}^n  w_\kappa(\zeta)  D_{i_\kappa}(\zeta)  \;  
\in {\cal  J}_{i_1\cdots i_n} \; ,
\eeq
for some $\omega_\kappa \in P[\zeta]$. Irrespective of the monomial order, a (reduced) Gr\"obner basis is $\mathcal{G} = \{g_1\} = \{ 1 \}$.  
Eq.~(\ref{Eq:DecGenI}) becomes
\beq
{\cal N}_{i_1\cdots i_n}(\zeta) = {\cal N}_{i_1\cdots i_n}(\zeta)  \times 1  \;  
\in {\cal  J}_{i_1\cdots i_n} \; ,
\eeq 
thus ${\cal I}_{i_1\cdots i_n}$   is reducible. 
\end{proof}

\subsection{Integrand Recursion Formula}
After substituting Eqs.~(\ref{Eq:DecGenI}) and (\ref{Eq:fromGtoD}) in  Eq.~(\ref{Eq:Igen}), we 
get a non-homogeneous recurrence  relation for the $n$-denominator integrand,
\beq
\mathcal{I}_{i_1\cdots i_n}  =  
 \sum_{\kappa=1}^{n}   \mathcal{I}_{i_1\cdots i_{\kappa -1} i_{\kappa+1} i_n}
+ \frac{\Delta_{i_1\cdots i_n}}{\db{i_1} \cdots  \db{i_n}}  .
\label{Eq:DecGen}
\eeq
According to Eq.~(\ref{Eq:DecGen}), 
$\mathcal{I}_{i_1\cdots i_n}$ is expressed in terms of integrands, 
$\mathcal{I}_{i_1\cdots i_{\kappa -1}  i_{\kappa+1} i_n}$, with 
$(n-1)$ denominators.
$\mathcal{I}_{i_1\cdots i_{\kappa -1}  i_{\kappa+1} i_n}$ are obtained 
from $\mathcal{I}_{i_1\cdots i_n}$
by pinching the  $i_\kappa$-th denominator.
The numerator of the non-homogeneous term is 
the remainder $\Delta_{i_1\cdots i_n}$ of the division~(\ref{Eq:DecGenI}). 
By construction, it contains only irreducible monomials 
with respect to ${\cal G}_{i_1\cdots i_n}$, thus it is 
identified with the {\it residue} at the cut $(i_1\ldots i_n)$.

The integrands $\mathcal{I}_{i_1\cdots i_{\kappa -1}  i_{\kappa+1}  \cdots i_n}$ can be decomposed
repeating the procedure described in  Eqs.~(\ref{Eq:Igen})-(\ref{Eq:DecGenI}). In this case the 
polynomial division of $\mathcal{N}_{i_1\cdots i_{\kappa -1}
  i_{\kappa+1}  \cdots i_n}$ has to be performed 
  modulo the Gr\"obner basis of the ideal 
${\cal J} _{i_1\cdots i_{\kappa -1}   i_{\kappa+1}  \cdots i_n}$, generated by 
the corresponding $(n-1)$ denominators.

 The complete multi-pole decomposition of the integrand $\mathcal{I}_{i_1\cdots i_n}$
 is achieved by successive iterations of 
Eqs.~(\ref{Eq:Igen})-(\ref{Eq:DecGenI}). 
Like an Erathostene's sieve, the recursive application of
Eqs.~(\ref{Eq:DecGenI}) and~(\ref{Eq:DecGen}) extracts the unique structures of the remainders 
$\Delta$'s. 
The procedure  naturally stops when all cuts are
exhausted, and no denominator is left, leaving us with the integrand
reduction formula.
 
If all quotients of the last divisions vanish, the integrand is  {\it cut-constructible}, i.e.
it can be  determined by sampling the numerator on the solutions of the cuts.  
If the quotients do not vanish, they give rise to 
{\it non-cut-constructible terms}, i.e. terms vanishing at every multi-pole. 
They can be reconstructed by sampling 
the numerator away from the cuts.
Non-cut-constructible terms may occur in
non-renormalizable theories, where the rank of the numerator is higher
than the number of denominators~\cite{Mastrolia:2012bu}.

\medskip

The  Proposition~\ref{Prop:Red4D} and the recurrence relation~(\ref{Eq:DecGen}) 
are the two mathematical properties underlying the integrand decomposition
of any scattering amplitudes. The polynomial form of each
residue is univocally derived from the division modulo the Gr\"obner
basis of the corresponding cut.

\section{One-loop integrand decomposition}
\label{Sec:OneLoop}

In this section we decompose an $n$-point integrand $\mathcal{I}_{0\cdots (n-1)}$ of rank-$n$ 
with $n>5$, using the procedure described in Section~\ref{Sec:Gen}. The reduction of 
higher-rank and/or lower-point integrands proceeds along the same lines.

\medskip

In $d$-dimensions, the   generic $n$-point one-loop integrand reads as follows: 
\bea
 \mathcal{I}_{0\cdots (n-1)} \equiv \frac{{\cal N}_{0\cdots(n-1)}(q,\mu^2)}{D_{0}(q,\mu^2) \cdots D_{n-1}(q,\mu^2)} \; .
 \label{Eq:Gamma0N}
 \eea
 We closely follow  the notation  of~\cite{Mastrolia:2010nb, Mastrolia:2012bu}. Objects living in $d=~4-2\epsilon$  are denoted by a bar, e.g. 
$\slh{{\bar q}} = \slh{q} + \slh{\mu}$ and   ${\bar q}^2= q^2 - \mu^2$.

\medskip

For later convenience, for each $\mathcal{I}_{i_1\cdots i_k}$ we define a 
basis $\mathcal{E}^{(i_1\cdots i_k)}= \{e_i\}_{i=1,\ldots,4}$. \\
If $k \ge 5$, then $e_i = k_i$, where $k_i$ are four external momenta. \\
If $k < 5$, then $e_i$ are chosen to fulfill the following relations: 
 \begin{align}
 & e_{1}^2 = e_2^2 = 0 \; ,    & & e_1\cdot e_2 =1\; ,  \nn
 & e_3^2 = e_4^2 = \delta_{k4} \; ,  & &     e_3 \cdot e_4 =-(1-\delta_{k4}) \; .
 \end{align}
In terms of $\mathcal{E}^{(i_1\cdots i_k)}$, the loop momentum can be decomposed as,
\beq
q^\mu  = -p^\mu_{i_1} +  x_1 \ e^\mu_1 + x_2 \ e^\mu_2 + x_3 \ e^\mu_3 + x_4 \ e^\mu_4  \; .
\label{eq:basi4}
\eeq
Accordingly, each  numerator ${\cal N}_{i_1 \cdots i_k}$ 
can be treated as a rank-$k$ polynomial in 
$\zeta\equiv (x_1,x_2,x_3,x_4, \mu^2)$,
\beq
{\cal N}_{i_1 \cdots i_k}  = \sum_{{\vec j \in J(k)}} \alpha_{\vec j} \; z_1^{\, j_1}  \, z_2^{\, j_2}  \; z_3^{\, j_3} \, z_4^{\, j_4} \, z_5^{\, j_5}  \; , 
\label{Eq:Ngeneral}
\eeq
with $J(k) \equiv \{\vec j = (j_1, \ldots, j_5) :  j_1+j_2+j_3+j_4+2\, j_5 \le k \}$. 

\medskip

{\it Step 1.}
When $n>5$, the Proposition~\ref{Prop:Red4D} guarantees that 
 ${\cal N}_{0\cdots n-1}$ is reducible,
and, by iteration, it can be written as a linear combination of  $5$-point integrands  $\mathcal{I}_{i_1 \cdots i_{5}}$.  

\medskip

{\it Step 2.}
 The numerator of each $ \mathcal{I}_{i_1\cdots i_5}$ is a rank-5 polynomial in $\zeta$, cfr. 
Eq.~(\ref{Eq:Ngeneral}).
We define the ideal ${\cal J}_{i_1\cdots i_5}$, and compute the
Gr\"obner basis ${\cal G}_{i_1\cdots i_5}=(g_1,\ldots,g_5)$, which is found to have
a remarkably simple form:
\beq
g_i(\zeta) = c_i + z_i \ , (i=1,\ldots,5) \ . 
\label{eq:GB5ple}
\eeq
We observe that each $g_i$ depends {\it linearly} on the $i$-th component of $\zeta$.

The division of ${\cal N}_{i_1 \cdots i_5}$ modulo
${\cal G}_{i_1\cdots i_5}$, see Eq.(\ref{Eq:DecGenI}), 
gives a {\it constant} remainder, 
\beq
\Delta_{i_1\cdots i_5} = c_0 \ .
\label{eq:Resi5}
\eeq
The term $\idea _{i_1\cdots i_5}$ in Eq.~(\ref{Eq:fromGtoD}) is,
\bea
\idea_{i_1\cdots i_5}  =  \sum_{\kappa=1}^{5}   
{\cal N}_{i_1\cdots i_{\kappa -1}i_{\kappa+1}\cdots i_5}(\zeta)
\db{i_\kappa}(\zeta) \; ,
\nonumber
\eea  
where ${\cal N}_{i_1\cdots i_{\kappa -1}i_{\kappa+1}\cdots i_5}$
are the numerators of the 4-point integrands, 
${\cal I}_{i_1\cdots  i_{\kappa -1}i_{\kappa+1}\cdots i_5}$,
obtained by removing the $i_\kappa$-th denominator.

\medskip

{\it Step 3.}
 For each $ \mathcal{I}_{i_1\cdots i_4}$, 
the  numerator   ${\cal N}_{i_1 \cdots i_4}$ is a rank-$4$ polynomial in $\zeta$.
The Gr\"obner basis ${\cal G}_{i_1\cdots i_4}$ of the ideal ${\cal J}_{i_1\cdots i_4}$
contains four elements.
Dividing ${\cal N}_{i_1 \cdots i_4}$ modulo 
${\cal G}_{i_1\cdots i_4}$, we obtain the remainder. The latter
depends  on $\mu^2$ and on the fourth component of 
the loop momentum $q$ in the basis 
$\mathcal{E}^{(i_1\cdots i_4)}$, 
\bea
\Delta_{i_1\cdots i_4} &=& c_0 + c_1 x_4 \nn
&+& \mu^2( c_2 + c_3 x_4 + \mu^2 c_4 ) \, . 
\eea
The term $\idea _{i_1\cdots i_4}$,
\bea
\idea _{i_1\cdots i_4}   =  \sum_{\kappa=1}^{4}   
{\cal N}_{i_1\cdots i_{\kappa -1}i_{\kappa+1}\cdots i_4}(\zeta)
\db{i_\kappa}(\zeta) \; ,
\nonumber
\eea  
contains the numerators of 3-point integrands 
${\cal I}_{i_1\cdots  i_{\kappa -1}i_{\kappa+1}\cdots i_4}$.

\medskip

{\it Step 4.}
The Gr\"obner basis
${\cal G}_{i_1 i_2 i_3}$ is formed by three elements, and is used to divide
${\cal N}_{i_1 i_2 i_3}$. The remainder ${\Delta}_{i_1 i_2 i_3}$ is
polynomial in $\mu^2$ and in the third and fourth components of $q$ 
in the basis ${\cal E}^{(i_1 i_2 i_3)}$,
\bea
{\Delta}_{i_1 i_2 i_3} &=& c_0 + c_1 x_3 + c_2 x_3^2 + c_3 x_3^3 \nn
&+& c_4 x_4 + c_5 x_4^2 + c_6 x_4^3 \nn
&+& \mu^2(c_7 + c_8 x_3 + c_9 x_4) \ .
\eea
The term $\idea_{i_1 i_2 i_3}$ generates
the rank-$2$ numerators of the $2$-point integrands 
${\cal I}_{i_1 i_2}$,  ${\cal I}_{i_1 i_3}$, and ${\cal I}_{i_2 i_3} $.

\medskip

{\it Step 5.}
The remainder of the division of ${\cal N}_{i_1 i_2}$ by the two elements of 
${\cal G}_{i_1 i_2}$ is:
\bea
{\Delta}_{i_1 i_2} &=& c_0  + c_1 x_2 + c_2 x_3 + c_3 x_4 \nn
&+&   c_4 x^2_2 + c_5 x^2_3 + c_6 x^2_4 + c_7 x_2 x_3 \nn
&+& c_9 x_2x_4  + c_9 \mu^2 \ . \quad 
\eea
It is polynomial in $\mu^2$ and in the last three components of $q$ in the basis 
$\mathcal{E}^{(i_1 i_2)}$.
The reducible term of the division, $\idea_{i_1 i_2} $,
generates the  rank-$1$  integrands, 
${\cal I}_{i_1}$, and ${\cal I}_{i_2}$.

\medskip

{\it Step 6.}
The  numerator of the 1-point integrands is linear in the components of the loop momentum 
in the basis $\mathcal{E}^{(i_1)}$,
  \beq
{\cal N}_{i_1}= \beta_0 +
\sum_{j = 1 }^4 \beta_{j} \;  x_{j} \ . \nonumber
\eeq
The only element of the Gr\"obner basis ${\cal G}_{i_1}$ is $D_{i_1}$, which is
quadratic in $\zeta$. Therefore the division modulo ${\cal G}_{i_1}$,
leads to a vanishing quotient, hence
\beq
{\cal N}_{i_1} = \Delta_{i_1} \ .
\eeq

\medskip

{\it Step 7.}
Collecting all the remainders computed in the previous steps, we obtain
the complete decomposition of $\mathcal{I}_{0\cdots n-1}$ in terms of its multi-pole structure
 \beq
 \mathcal{I}_{0\cdots n-1}  =  \sum_{k=1}^5 \; \left (  
\sum_{1=i_1 < \ldots < i_k}^{n-1}  \; \; 
\frac{\Delta_{i_1 \cdots i_k}  }{D_{i_1} \cdots D_{i_k}   } \right ) \, .
\label{Eq:Reduction}
 \eeq
 Eq.~(\ref{Eq:Reduction})
reproduces the well-known one-loop $d$-dimensional integrand decomposition formula 
\cite{Ossola:2006us,Ellis:2007br,Giele:2008ve,Ellis:2008ir,Mastrolia:2010nb,Ellis:2011cr}.

We remark that
the basis $\mathcal{E}^{(i_1\cdots i_k)} $, defined in
Eq.(\ref{eq:basi4}) and used for decomposing the integration momentum $q$,
depends only on the external momenta of diagram associate to the cut,
eventually complemented by orthogonal elements.
Therefore, $\mathcal{E}^{(i_1\cdots i_k)} $ can be used as well to
decompose the integration momenta of multi-loop diagrams \cite{Mastrolia:2011pr}.

\section{The Maximum-cut Theorem}
\label{Sec:MaximumCut}

At $\ell$ loops, in four dimensions, we define a {\it maximum-cut}  
as a $(4 \ell)$-ple cut
\beq
D_{i_1}=D_{i_2}=\cdots = D_{i_{4\ell}} =0 \; ,
\label{Eq:Maximum}
\eeq
which  constrains completely the components of the loop
momenta. In four dimensions this implies the presence of four 
constraints for each loop momenta.  
We assume that, in non-exceptional phase-space points, a maximum-cut has
a finite number $n_s$ of
solutions, each with multiplicity one. Under this assumption we have the following

\begin{theo}[Maximum cut]
The residue at the maximum-cut is a polynomial paramatrized by 
$n_s$ coefficients, which admits a univariate representation of degree $(n_s-1)$.
\label{PropM}
\end{theo}
%
\begin{proof}
Let us parametrize the propagators using $4 \ell$
variables $\zeta = (z_1, \ldots, z_{4 \ell})$. In this parametrization,
the solutions of the maximum-cut read,
\beq
\zeta^{(i)}  = \left (z^{(i)}_1, \ldots , z^{(i)}_{4 \ell}  \right ) \ , \; {\rm with} \,  i=1,\ldots,n_s \ .
\eeq

Let $\mathcal{J}_{i_1\cdots i_{4\ell}}$ be the ideal generated by the on-shell denominators,
$
\mathcal{J}_{i_1\cdots i_{4\ell}} = \langle D_{i_1}, \ldots, D_{i_{4\ell}}\rangle \ .
$ \\
According to the assumptions, the number $n_s$ of the solutions 
of~(\ref{Eq:Maximum}) is finite, 
and each of them has multiplicity one, therefore 
 $\mathcal{J}_{i_1\cdots i_{4\ell}}$ is zero-dimensional \cite{moller}
and radical\footnote{
Given an ideal ${\cal J}$, the {\it radical} of ${\cal J}$ is  
$\sqrt{{\cal J}} \equiv \{ f \in P[\zeta]: \ \exists \; s \in {\mathbb N}, \ f^s \in {\cal J} \}$. 
${\cal J}$ is radical iff ${\cal J} = \sqrt{{\cal J}}$.
}, see Cor. 2.6, Ch. 4 of \cite{Cox2}.
In this case, the {\it Finiteness Theorem}  
(Prop. 8, Ch. 5 of \cite{Cox1})
ensures that
the remainder of the division of any polynomial modulo $\mathcal{J}_{i_1\cdots i_{4\ell}}$
can be parametrized exactly by $n_s$ coefficients.

\medskip

Moreover,
up to a linear coordinate change,
we can assume that all the solutions of the system 
have distinct first coordinate $z_1$, i.e.
$z^{(i)}_{1} \neq z^{(j)}_{1}$  $\forall\; i \neq j$. 
We observe that $\mathcal{J}_{i_1\cdots i_{4\ell}}$ and $z_1$ are in the 
{\it Shape Lemma} position 
(Prop. 2.3 of \cite{sturmfels})
therefore
a  Gr\"obner basis for the lexicographic order $z_1<z_2<\cdots < z_n$ is 
$\mathcal{G}_{i_1\cdots i_{4\ell}} =\{ g_1, \ldots, g_{4\ell} \}$, in the form
\begin{equation} \label{eq:ShapeLemmaG}
\left\{
\begin{array}{ccc}
g_1(\zeta) &=& f_1(z_1) \\
g_2(\zeta) &=& z_2 - f_2(z_1) \\
&\vdots& \\
g_{4\ell}(\zeta) &=& z_{4\ell} - f_{4\ell}(z_1) \; .
\end{array}  \right.
\end{equation}
The functions $f_i$ are univariate polynomials in $z_1$. In particular
$f_1$ is a rank-$n_s$ square-free polynomial \cite{rouillier}, 
\beq
f_1(z_1) = \prod_{i=1}^{n_s} \; \left  (z_1 - z_1^{(i)} \right ) \ ,
\eeq
i.e. it does not exhibits repeated roots.
The multivariate division of ${\cal N}_{i_1\cdots\i_{4\ell}}$ modulo $\mathcal{G}_{i_1 \cdots i_{4\ell}}$
leaves a remainder $\Delta_{i_1\cdots i_{4\ell}}$ which is a univariate polynomial in $z_1$ of degree
$(n_s-1)$ \cite{verschelde}, in accordance with the {\it Finiteness Theorem}.
\end{proof}

\medskip

The  maximum-cut theorem  ensures that the maximum-cut residue, at
any loop, is completely determined by the $n_s$ distinct solutions of
the cut-conditions. In particular it can be reconstructed by sampling
the integrand on the solutions of the maximum cut itself.

At one loop and in $(4-2 \epsilon)$-dimensions, the $5$-ple cuts are maximum-cuts. The remarkably simple 
structure of the Gr\"obner basis in Eq.~(\ref{eq:GB5ple}) 
is dictated by the maximum-cut theorem.
Moreover in this case $n_s=1$, thus the residue in Eq.~(\ref{eq:Resi5}) is a constant.

\begin{figure}
\begin{center}
\includegraphics[width=0.47\textwidth]{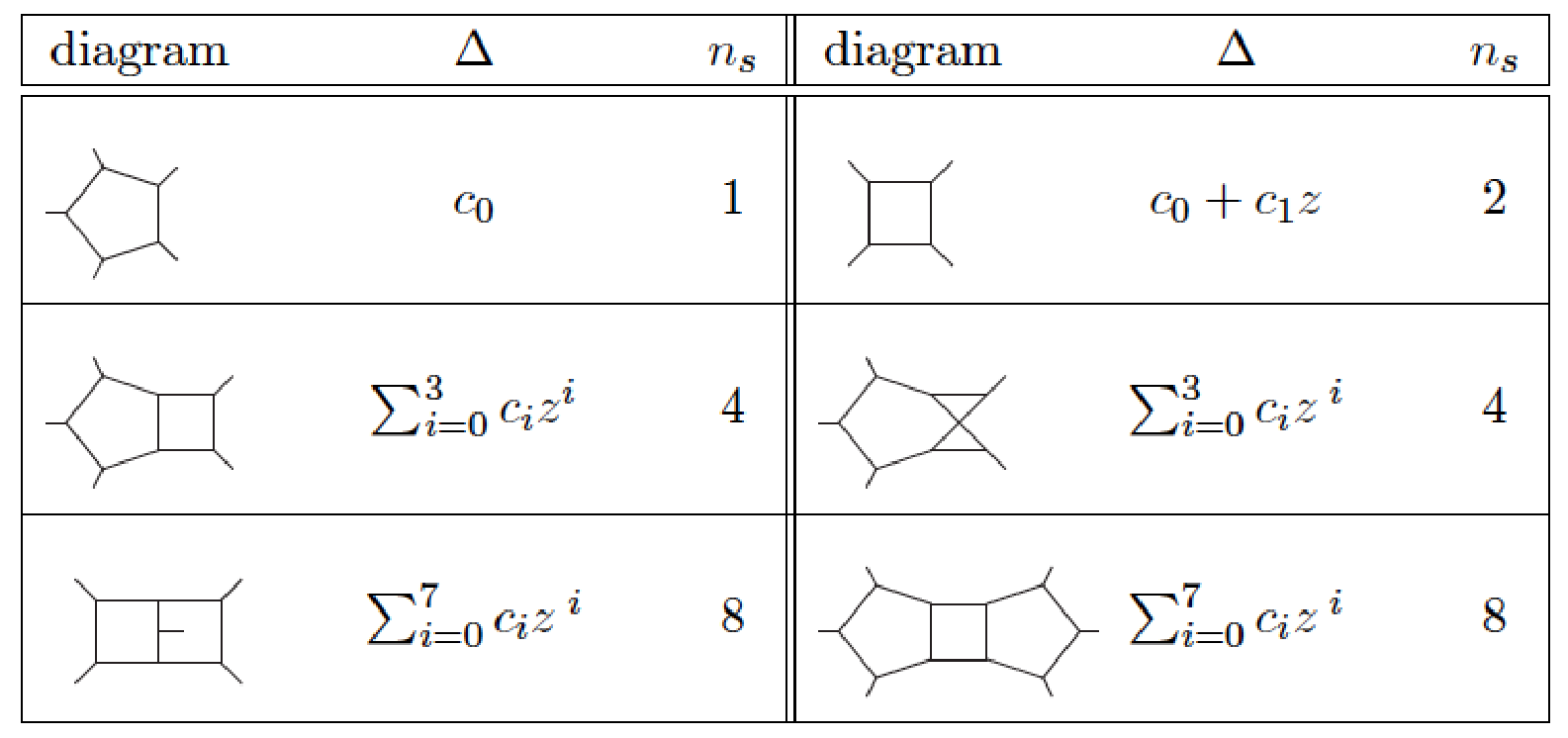}
\end{center}
\caption{
The on-shell diagrams in the picture are examples of maximum-cuts.
The first diagram in the left column represents the 5-ple cut of the 5-point 
one-loop dimensionally regulated amplitude. All the other on-shell diagrams are considered 
in four dimensions.
For each of them, the general structure of the residue $\Delta$ (according to the Shape Lemma) 
and the corresponding  value of $n_s$ are provided.}
\label{tableMC}
\end{figure}

The structures of the residues of the maximum cut, together with the corresponding 
values of $n_s$, for a set of one-, two-, and three-loop diagrams 
are collected in Figure~\ref{tableMC}.

\medskip

The calculations of Sections \ref{Sec:OneLoop} and \ref{Sec:MaximumCut}
have been carried out using the package {\sc S@M} \cite{Maitre:2007jq}
and the functions {\tt GroebnerBasis} and {\tt PolynomialReduce} of
{\sc Mathematica},
respectively needed for the generation of the Gr\"obner basis and the polynomial
division.

\section{Conclusions}
We presented a new algebraic approach, where the evaluation of scattering amplitudes 
is addressed by using multivariate polynomial division, with
the components of the loop-momenta as indeterminates.
We found a recurrence relation to construct the integrand decomposition 
of arbitrary scattering amplitudes, independently 
of the number of loops. 
The recursive algorithm is based on 
the Weak Nullstellensatz Theorem
and on the division modulo the Gr\"obner basis associated to all
possible multi-particle cuts.
Using this technique, 
we rederived the well-known one-loop integrand decomposition formula.
Finally, by means of the Finiteness Theorem and of the Shape Lemma, 
we proved that the residue at the maximum-cuts is parametrised 
exactly by a number of coefficients equal to the number of solutions of the cut itself.

\section*{Acknowledgments}
We are indebted to Simon Badger and Yang Zhang for fruitful discussions, 
in particular on the properties of the Gr\"obner basis, and for comments of the manuscript.

E.M.  thanks the Center for Theoretical Physics of New York City College of
Technology for hospitality during the final stages of this project.

The work of P.M. and T.P. is 
supported by the Alexander von Humboldt Foundation, in the framework of the Sofja Kovaleskaja
Award, endowed by the German Federal Ministry of Education and Research.
The work of G.O. is  supported in part by the National Science Foundation under Grant PHY-1068550. 




\bigskip
\bigskip

\bibliographystyle{utphys} 

\bibliography{references}

\end{document}